# A Study on Impact of Dividend Policy on Initial Public Offering Price Performance

¹S. Meghna, ²N. Suresh and ²J.C. Usha
¹Department of MBA-Financial Management,
²Faculty of Management and Commerce, Ramaiah University of Applied Sciences,
560054 Bangalore, India

**Abstract:** This study examines the impact of dividend policy on the performance of initial public offering in India. The period of study is from the year 2011-2014. Monthly returns of the IPOs issued in the considered period and Indian Stock Market Index (Nifty 50) was considered for the long-run performance study. The methodological tools used are long-run performance statistics and garch model. 'Dummy' variable was used to measure the effect of dividends on the IPOs. The study reveals that the dividend policy has no significant effect on the stock prices of IPO.

**Key words:** IPO, long-run, wealth relatives, dividends, effect, stock prices

## INTRODUCTION

An Initial Public Offering (IPO) is the initial offer of shares by a privately owned organization to the general public and is a vital method for fund-raising from public by organizations. The initial shares are sold to institutional investors and also to the retail investors. One or more investment banks help in underwriting of IPOs, who also arrange for the shares to be listed on one or more stock exchange. Through this process, known as 'floating' or 'going public', a private organization is transformed into a public organization. A public organization has access to more and often deeper sources of capital than a private organization. In this study, we examine the effect of dividend on stock price of IPOs for first 35 months of their issue. A payment made by an organization to its shareholders which is usually, a distribution of profits is called issuing a dividend. When an organization earns a profit they re-invest the profit in the business and a proportion of the profit is paid as a dividend to the organization's shareholders . Distribution of dividend to shareholders may be in the form of cash or the amount can be paid by the issue of further shares or share repurchase, if the corporation has a dividend reinvestment plan.

**Literature review:** There has been a lot of research on various aspects of IPOs. Krishnamurti and Kumar (2002) studied on the price performance at the time of listing and confirmed the phenomenon of under-pricing of IPO with the short run analysis of Indian IPOs issued from 1992-1994. Bharat A. Jain, Chander Shekar and Violet Torbey studied Australian market focusing on the initiation factors of dividend on new public firms issued from 1990-1998 by concluding that dividend issuing firms are stable. G. Sabarinathan talks about different Indian IPO events since the establishment of the Securities and Exchange Board of India (SEBI) in the year 1992. Fernando et al. examined the value execution of Initial Public Offerings (IPOs) in India among 27 book building IPOs in India including a time of 5 years from 1990-2004 and demonstrated that there is no much distinction amongst low and high premium issues and proposed that for the most part low premium issues are under-valued and are more predictable as respects to returns than that of the high premium issues. Dhamija and Arora (2014) focused on long-run performance of Indian IPO firms among different grades awarded issued from 2007-2010 and the results imply that the long-run performance of Indian IPOs does not bear any relationship with the grade awarded. Asrin Sheikh, Ahmadian, Parastoo Sedaghat and Manasa and Narayanarao (2018) studied the long-run and short-run performance analysis of Indian IPOs issued from 2007-2012 and concluded that there is a significant relationship between IPOS, market timing and capital structure and the IPOs have short-term effects on capital structure of the firm. Handa and Singh (2017) focused on the underpricing phenomenon of Indian IPO firms issued from 2001-2012 where the rersearchers concluded that corporate governance mechanisms like BoD, Audits, signaling, underwriters etc., influence the underpricing phenomenon at the time of firm going public.

**Corresponding Author:** N. Suresh, Faculty of Management and Commerce, Ramaiah University of Applied Sciences,
560054 Bangalore, India

From the above literature review it can be concluded that there is scope for study on effect of dividend policy on Indian IPOs on its long run performance. The focus of this study is to compare the long run performance of Indian IPOs among different grades (1-Grade5) and to examine the effect of dividend policy on the considered IPOs issued from the year 2011-2014 (Agarwal *et al.*, 2008; Al-Shawawreh and Al-Tarawneh, 2015).

## MATERIALS AND METHODS

Graded IPOs listed on the National Stock Exchange (NSE) were considered for the purpose of this study. Graded IPOs that have completed at least 3 years, since, the date of their listing are only considered. The grading is done on a five point scale as given below: IPO Grade 1-Poor fundamentals, IPO Grade 2-Below-average fundamentals, IPO Grade 3-Average fundamentals, IPO Grade 4-Above-average fundamentals, IPO Grade 5-Strong fundamentals.

**Long-run performance test:** The sample that is considered for the long-run performance study comprises of 39 IPOs made by Indian companies during the period 2011-2014 which have been graded by credit-rating agencies. Monthly closing prices of the sample companies on NSE for a period of 36 months from the date of listing were taken. Methodology used to evaluate long-run performance of IPOs by Ritter and Welch (2002) is as follows:

- Cumulative average raw returns calculated with monthly portfolio rebalancing
- Cumulative average adjusted returns calculated with monthly portfolio rebalancing where the adjusted returns are computed using returns of the index (Nifty50) of the National Stock Exchange as the benchmark.
- The 3-year buys and holds returns
- Wealth relatives

The initial return period is defined to be 1 month and the after market period includes the following 36 months where a month is defined as a period of successive 21-trading-day period relative to the IPO date. Thus, month 1 consists of event days 2-22; month 2 consists of event days 23-43, etc. Monthly benchmark-adjusted returns are calculated as the monthly raw returns on a stock minus the monthly benchmark returns for the corresponding 21-trading-day period. The benchmark-adjusted returns ($ar_{it}$) for stock i in event month t is defined as Eq. 1:

$$ar_{it} = r_{it} - r_{mt} \qquad (1)$$

Where:
$r_{it}$ = The returns on security i in the event month t and $r_{mt}$ = The benchmark returns in event month t

The average benchmark-Adjusted Returns ($AR_t$) on a portfolio of n stocks for event month t are the equally-weighted arithmetic average of the benchmark-adjusted returns Eq. 2:

$$AR_t = 1/n \sum ar_{it} \qquad (2)$$

The cumulative benchmark-adjusted aftermarket performance from event month q to event month s represents the sum of the average benchmark-adjusted returns Eq. 3:

$$CAR_{1,t} = \sum_{t=1}^{5} AR_t \qquad (3)$$

To interpret this 3-year total return, Wealth Relatives (WRs) can be calculated as a performance measure. WR is defined as: Wealth Relatives (WR) Eq. 4:

$$WR = \frac{(1+\text{Avg.3 year total returns on IPOs})}{(1+\text{Avg.3 year total returns on benchmark})} \qquad (4)$$

A wealth relative of >1.00 can be interpreted as IPOs outperforming a portfolio of matching firms; a wealth relative of <1.00 indicates that IPOs underperformed (Table 1).

**Dividend effect analysis:** The sample that is considered for the effect of dividends on IPOs study comprises of 27 out of 39 IPOs made by Indian companies considered for the long-run performance as the companies like Grade1 (4 IPOs)-Onelife Capital Advisors Limited, Timbor Home limited, Shilpi Cable Technologies Limited, VKS projects limited, Grade2 (5 IPOs)-Indo Thai Securities Limited, Prakash Constrowell Ltd., Brooks Laboratories Limited, Vaswani Industries Limited, Servalakshmi Paper Limited, Grade3 (3 IPOs)-PG Electroplast Limited, Sanghvi Forging and Engineering Ltd, Tara Jewels Limited have not issued the dividends. Monthly closing prices of the sample companies on NSE for a period of 36 months from the date of listing were considered (Brau *et al.*, 2003; Carter and Manaster, 1990).

Table 1: A wealth relative of <1.00 indicates that IPOs underperformed

| Grades | Firms names |
|---|---|
| Grade 1 (6 IPOs) | Onelife Capital Advisors Limited, Timbor Home Limited, Shilpi Cable Technologies Limited, VKS projects Limited, Shemaroo Entertainment Limited, Sharda Cropchem Limited |
| Grade 2 (7 IPOs) | Indo Thai Securities Limited, Prakash Constrowell Ltd., Brooks Laboratories Limited, Inventure Growth and Securities Ltd, Rushil Decor Limited, Vaswani Industries Limited, Servalakshmi Paper Limited |
| Grade 3 (14 IPOs) | Repco Home Finance Limited, V-mart Retail Limited, Flexituff International Limited, PG Electroplast Limited, SRS Limited, Tree House Education and Accessories Limited, Sanghvi Forging and Engineering Ltd, Lovable Lingerie Limited, Acropetal Technologies Limited, Omkar Speciality Chemicals Limited, PC Jeweller Limited, Tara Jewels Limited, Tribhovandas Bhimji Zaveri Limited, Innoventive Industries Limited |
| Grade 4 (9 IPOs) | TD Power Systems Limited, Muthoot Finance Limited, PTC India Financial Services Limited, Bharti Infratel Limited, Speciality Restaurants Limited, MT Educare Limited, Monte Carlo Fashions Limited, Snowman Logistics Limited, Wonderla Holidays Limited |
| Grade 5 (3 IPOs) | Just Dial Limited, L and T Finance Holdings Limited, Multi Commodity Exchange of India Limited |

GARCH (1, 1) Model is developed to find the effect of dividends on IPOs. The development of GARCH (1, 1) Model consists of two equations viz., mean equation and variance equation. Mean equation is represented by:

$$\text{Ind.Stock} = C1 + C2 * \text{NIFTY50} + e \quad (1)$$

And variance Eq. 2 is represented by:

$$\text{GARCH} = C3 + C4 * (\text{RESID}(-1))^2 \\ C5 * \text{GARCH}(-1) + C6 * \text{DUMMY} + e \quad (2)$$

GARCH = Residual variance (error term) derived from the mean Eq. 1. It is also known as present/current day's variance or volatility of Stock Market Return (Nifty) (Reber and Fong, 2011).

$(\text{RESID}(-1))^2$ = Previous periods' residual square obtained from Eq. 1 and is known as Lag/previous day's return infor mation regarding volatility. It is known as ARCH term.

GARCH (-1) = Lag/Previous day's variance residual or volatility of Stock Market Return (Nifty 50). Term is known as GARCH. DUMMY = Variable to represent period of dividend paid to calculate its effect on the IPOs.

## RESULTS AND DISCUSSION

**Raw returns:** Table A1-A4 (refer Annexure A) report Average Raw returns (ARt) and Cumulative Average Returns (CAR1, t) of IPOs made during the period 2011-2014. The returns are calculated in per cent for 36 months after going public, excluding the initial return. Figure 1 shown below trend line defines the cumulative raw returns for different grades from 1-5 (Jaitly, 2004; Loughran and McDonald, 2013).

For IPOs with grade 1, 34 of the 36 monthly average raw returns are negative. Cumulative average raw returns are -20.73% by the end of month 36, exclusive of the initial returns. For IPOs with grade 2, 33 of the 36 monthly average raw returns are negative. Cumulative average raw returns are 38.49% by the end of month 36, exclusive of the initial returns. For IPOs with grade 3, 8 of the 36 monthly average raw returns are negative. Cumulative average raw returns are 19.83% by the end of month 36, exclusive of the initial returns. For IPOs with grade 4, 15 of the 36 monthly average raw returns are negative. Cumulative average raw returns are 5.33% by the end of month 36, exclusive of the initial returns. For IPOs with grade 5, 2 of the 36 monthly average raw returns are negative. Cumulative average raw returns are 38.05% by the end of month 36, exclusive of the initial returns. Figure 2 and Table 2 summarize the findings on the monthly average and cumulative average raw returns.

**Benchmark adjusted returns:** Average benchmark-Adjusted Returns (Art) and CAR1,t of IPOs issued in the period 2011-2014 in percent for 36 months after going public, excluding the initial return. Figure 3 shown below provide trend line defines the cumulative benchmark adjusted returns of different grades from 1-5.

For IPOs with grade 1, 35 of the 36 monthly average adjusted returns are negative. Cumulative average adjusted returns are -34.94% by the end of month 36, exclusive of the initial return. For IPOs with grade 2, 35 of the 36 monthly average adjusted returns are negative. Cumulative average adjusted returns are .41.81% by the end of month 36, exclusive of the initial return. For IPOs with grade 3, 17 of the 36 monthly average adjusted returns are negative. Cumulative average adjusted returns are 2.08% by the end of month 36, exclusive of the initial return. For IPOs with grade 4, 33 of the 36 monthly average adjusted returns are negative. Cumulative average adjusted returns are 6.26% by the end of month 36, exclusive of the initial return. For IPOs with grade 5, 2 of the 36 monthly average adjusted returns are negative. Cumulative average adjusted returns are 17.53% by the end of month 36, exclusive of the initial return.

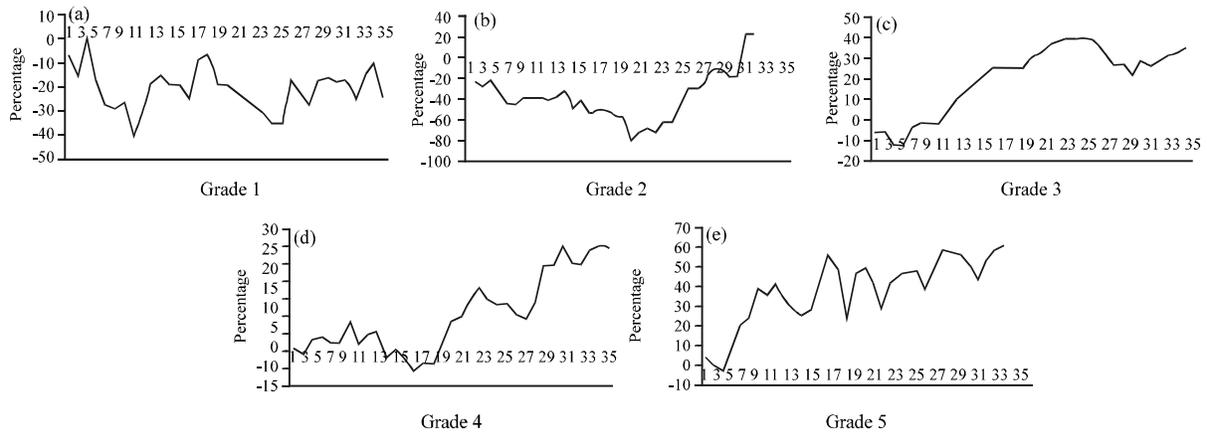

Fig. 1: a-e)Cumulative raw returns for different grades from 1-5

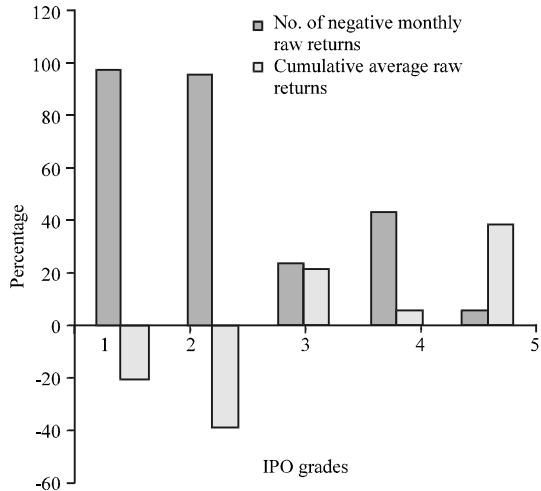

Fig. 2: Monthly average and cumulative average raw returns

Table 2: Monthly average and cumulative average raw returns

| No. of negative monthly raw returns | Cumulative average raw returns (%) |
|---|---|
| 34 | -20.73 |
| 33 | -38.49 |
| 8 | 19.83 |
| 15 | 5.33 |
| 2 | 38.05 |

Table 3: Monthly average and cumulative average

| No. of negative monthly benchmark adjusted returns | Cumulative average benchmark adjusted returns (%) |
|---|---|
| 35 | -34.94 |
| 35 | -41.81 |
| 17 | 2.08 |
| 33 | -6.26 |
| 2 | 17.53 |

Figure 4 and Table 2 summarize the findings on the monthly average and cumulative average benchmark-adjusted returns.

Table 4: The 3-year holding period returns of IPOs issued

| Grades | Grade 1 (%) | Grade 2 (%) | Grade 3 (%) | Grade 4 (%) | Grade 5 (%) |
|---|---|---|---|---|---|
| High | 113.60 | -55.80 | 133.43 | 102.15 | 74.03 |
| Low | -99.98 | -99.55 | -98.81 | -76.42 | -54.09 |
| Mean | -0.17 | -81.16 | 22.20 | -9.75 | 24.33 |
| Median | -4.04 | -85.94 | 68.05 | -43.87 | 53.05 |

Table 5: Report the WRs for different grades of IPOs

| Grades | Wealth relatives (%) |
|---|---|
| 1 | 57.89 |
| 2 | 10.37 |
| 3 | 70.11 |
| 4 | 53.42 |
| 5 | 68.50 |

**Holding period return:** Figure 5 and Table 4 show a distribution of holding period returns for different grades of IPO.

**Wealth relatives:** Figure 4 and Table 4 report the WRs for different grades of IPO. Grade 2 has performed the least, whereas the IPOs of grade 3-5 have performed well over the 3 consecutive years from the time of their issue.

**Dividend analysis:** 'Dummy' variable is used to measure the effect of dividends on the IPOs. Table 5 gives the results of the GARCH (1, 1) Model regression equation for IPOs of 1-grade 5. Shemaroo Entertainment Limited, Tree House Education and accessories Limited, V-mart Retail Limited, Monte Carlo Fashions Limited, Wonderla Holidays Limited and Just Dial Limited are significant at level 5%.

**Findings**
**Short run performance:** From the sample, 24 IPOs have given negative (initial) first day returns by defining the over-pricing phenomenon. The 38 IPOs

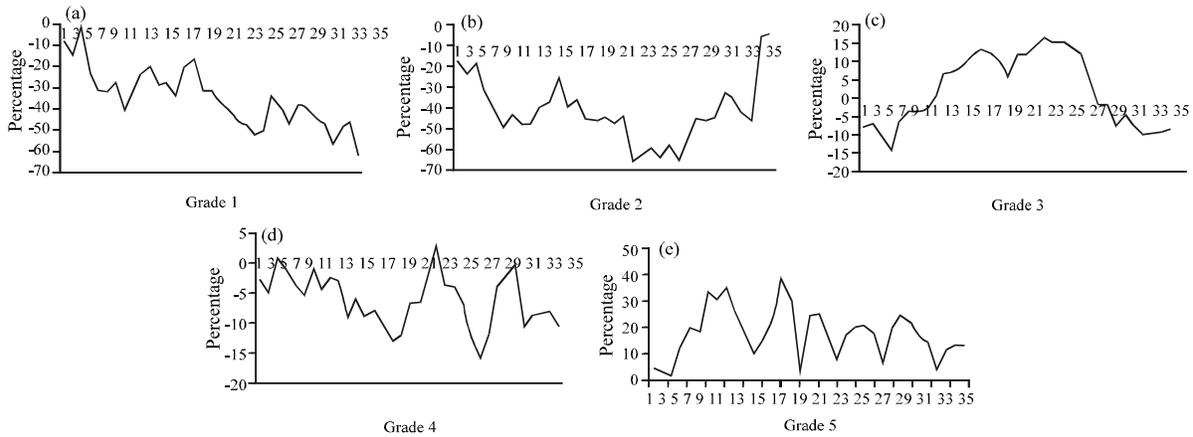

Fig. 3: a-e) Cumulative benchmark adjusted returns of different grades from 1-5

Table 6: GARCH (1, 1) Model regression equation for IPOs

| Companies | Mean equation | | Variance equation | | | |
|---|---|---|---|---|---|---|
| | c1 | Index | c3 | RESD(-1)$^2$ | GARCH | Dummy |
| Sharda Cropchem Limited | -0.0037(0.0100) | 1.4976*(0.2481) | 0.0039(0.0020) | -0.0906(0.0770) | 0.5685(0.3637) | -0.0041(0.0011) |
| Shemaroo Entertainment limited | 0.0071(0.0145) | 0.0972(0.4687) | 0.0027*(0.0009) | -0.1015*(0.0036) | 0.9685*(0.0151) | -0.0041*(0.0021) |
| Inventure Growth and Securities Ltd | 0.0026*(0.0062) | 0.1754*(0.3076) | 0.0002(0.0003) | 0.2906(0.3792) | 0.4557(0.4219) | 0.0155(0.0274) |
| Rushil Decor Limited | 0.0009(0.0391) | 0.8432(0.8787) | 0.0603*(0.0240) | 0.3364(0.2088) | -0.5120(0.2958) | -0.0213(0.0114) |
| Acropetal Technologies Limited | 0.0912*(0.0296) | 0.7912(0.7032) | 0.0052*(0.0006) | -0.3619*(0.0204) | 1.1380*(0.0186) | -0.0012(0.0009) |
| Flexituff International Limited | 0.0092(0.0096) | -0.0242(0.2100) | -0.0001(8.01E-05) | 0.0118(0.0428) | 0.7775*(0.0400) | 0.0010(0.0006) |
| Innoventive Industries Limited | 0.0201*(0.0101) | 0.4495*(0.1349) | 0.0015(0.0018) | 0.8592*(0.3501) | 0.4050*(0.1951) | -0.0029(0.0019) |
| Lovable Lingerie Limited | -0.0030(0.0250) | 0.6496(0.4642) | 0.0057(0.0043) | 0.0338(0.2144) | 0.5556(0.3893) | -0.0064(0.0040) |
| Omkar Speciality Chemicals Limited | 0.0202(0.0208) | 0.9069(0.5240) | 0.0011(0.0021) | -0.2602*(0.0007) | 1.1492*(0.0013) | 0.0048(0.0054) |
| PC Jeweller Limited | 0.0227(0.0264) | 1.4531*(0.5605) | 0.0022(0.0018) | -0.2254*(0.0998) | 1.1460*(0.0087) | -0.0006(0.0008) |
| Repco Home Finance Limited | 0.0188(0.0125) | 1.1071*(0.2703) | 0.0001(0.0007) | -0.3144*(0.0099) | 1.1994*(0.0001) | 0.0013(0.0016) |
| SRS Limited | -0.0205*(0.0091) | 0.9487*(0.2258) | 0.0059(0.0031) | -0.1943*(0.5185) | 0.5837(0.3566) | -0.0002(0.0034) |
| Tribhovandas Bhimji Zaveri Limited | -0.0214(0.0111) | 1.8216*(0.3020) | 0.0083(0.0071) | 1.1747*(0.5629) | -0.0625(0.1125) | -0.0074(0.0052) |
| Tree House Education and Accessories Limited | 0.0306*(0.0105) | 1.7149*(0.2530) | 0.0038*(0.0019) | -0.1107*(0.0305) | 0.4813*(0.1868) | -0.0040*(0.0018) |
| V-mart Retail Limited | -0.0142(0.0321) | 1.3833*(0.5692) | 0.0107(0.0093) | -0.1360(0.0815) | 0.6080(0.3245) | 0.0089*(8.83E-06) |
| Bharti Infratel Limited | 0.0095(0.0057) | 0.4255(0.4003) | 0.0026*(0.0012) | -0.0586(0.0453) | 0.6476*(0.1381) | -0.0027(0.0022) |
| Monte Carlo fashions Limited | 0.0078(0.0147) | 0.8514*(0.3297) | 0.0044(0.0034) | -0.0756(0.0462) | 0.7956*(0.2381) | -0.0051*(0.0019) |
| MT Educare Limited | -8.15E-05(0.0204) | 0.4910(0.4737) | 0.0319*(0.3441) | 0.3344(0.3441) | -0.3187(0.3410) | -0.0079(0.0048) |
| Muthoot Finance Limited | -0.0152(0.0259) | 1.3754*(0.5377) | -0.0005(0.0001) | -0.2480*(0.0048) | 1.2936*(0.0011) | 0.0055(0.0029) |
| PTC India Financial Services Limited | 0.0111(0.0059) | 0.2397*(0.0709) | 0.0006(0.0012) | -0.0316(0.2408) | 0.5481(0.9510) | -0.0007(0.0007) |
| Snowman Logistics Limited | -0.0192(0.0176) | 1.9830*(0.3617) | 0.0013(0.0016) | -0.3407*(0.1549) | 1.2230*(0.0012) | -0.0049(0.0102) |
| Speciality Restaurants Limited | -0.0270*(0.0111) | 1.2663*(0.2582) | 0.0048*(0.0024) | -0.2700*(0.1281) | 0.5915(0.5274) | -0.0020(0.0021) |
| TD Power Systems Limited | -0.0009(0.0166) | 0.3527(0.3197) | 0.0041(0.0051) | -0.1536*(0.0491) | 0.5824(0.6700) | -0.0007(0.0031) |
| Wonderla Holidays Limited | -0.0085(0.0126) | 0.2844(0.2328) | -0.0001(0.0003) | -0.0493*(0.0123) | 0.7899*(0.0835) | 0.0062*(0.0026) |
| Just Dial Limited | 0.0124(0.0270) | 1.0483*(0.4210) | 0.0130(0.0167) | -0.1634(0.1011) | 0.5959(0.6880) | -0.0257*(0.0114) |
| L andT Finance Holdings Limited | -0.0484(0.0265) | 1.9216*(0.6624) | 0.0056(0.0058) | -0.0544(0.0324) | 0.8438*(0.2049) | -0.0090(0.0054) |
| Multi Commodity Exchange of India limited | -0.0370(0.0243) | 1.9806 *(0.7101) | 0.0129 *(0.0060) | 0.8491(0.5069) | -0.0475(0.0399) | 0.0001(0.0112) |

*Indicates statistical significance at 5% level

from the sample considered have proved the under-pricing phenomenon by giving positive returns on the first day.

**Long run performance:** From the cumulative average raw returns and cumulative average benchmark adjusted returns it can be concluded that the IPOs with greater grades perform better when compared to the lower grades in a long run by considering the 3-year holding period returns of IPOs Issued, the greater graded IPOs have given better returns than that of lower grades.

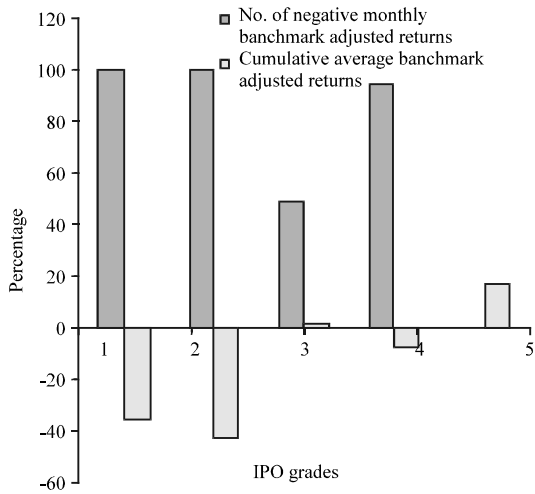

Fig. 4: Monthly average and cumulative average benchmark-adjusted

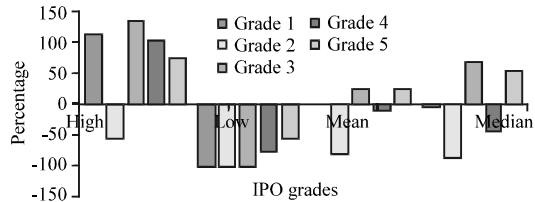

Fig. 5: A distribution of holding period returns for different grades of IPO

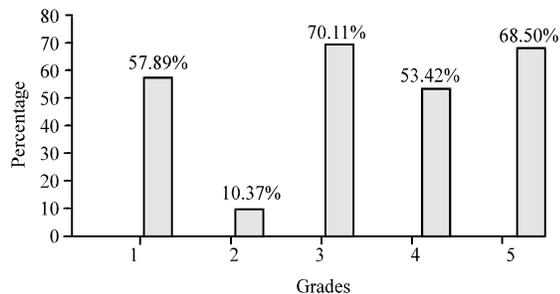

Fig. 6: Wealth relatives for different grades of IPO

**GARCH (1, 1) test for effect on dividend:** The 6 out of 27 IPOs issuing the dividend has an effect on the stock price which reveals that dividend does not have a significant influence on stock price.

## CONCLUSION

From the above observations the conclusion can be drawn as follows in short run performance from the sample, 24 IPOs have given negative (initial) first day returns by defining the over-pricing phenomenon and 38 IPOs from the sample considered have proved the under-pricing phenomenon by giving positive returns on the first day. Since, the number of observations of under-pricing phenomenon is larger that is 38 of 62 IPOs, it signifies the under-pricing phenomenon. From the cumulative average raw returns and cumulative average benchmark adjusted returns it can be concluded that the IPOs with superior grades perform better when compared to the lower grades in a long run and by considering the 3-year holding period returns of IPOs Issued, the greater graded IPOs have given better returns than that of lower grades.

Finally, the effect on the dummy column shows 6 out of 27 IPOs issuing the dividend has an effect on the stock price which reveals that dividend does not have a significant influence on stock price.

## LIMITATION

The study is limited to the Indian IPOs issued from the year 2011-2015 registered in NSE website. The limitation of this study is that the short run performance analysis is done considering the variance in the offer price of issued IPO.

## RECOMMENDATION

Future studies can be conducted on short run performance of IPOs with respect to the subscription level. The subscription level is the number of stocks issued by the IPO that are bought by the bidders.